\documentclass[a4paper]{article}
\usepackage{amssymb,amsmath}
\usepackage{chngcntr}
\usepackage{color}
\usepackage[latin1]{inputenc}
\usepackage[ruled,vlined,linesnumbered]{algorithm2e}
\usepackage[nice]{nicefrac}
\usepackage[ruled]{algorithm2e}
\usepackage{graphicx}
\usepackage{subfigure}
\usepackage{url,hyperref}
\usepackage{rotating}
\usepackage{multirow}

\begin{document}
\title{How do Ontology Mappings Change\\ in the Life Sciences? }

% wenn oben \documentclass[10pt]{llncs}
%\author{Anika Gross\inst{1,2} \and Michael Hartung\inst{1,2} \and Andreas Thor\inst{1} \and Erhard Rahm\inst{1,2}}
%\institute{
%Department of Computer Science, University of Leipzig \\
%\and Interdisciplinary Center for Bioinformatics, University of Leipzig \\ 
%\email{\{gross,hartung,thor,rahm\}@informatik.uni-leipzig.de}
%}
% wenn oben \documentclass[a4paper]{article}
\author{Anika Gross, Michael Hartung, Andreas Thor, Erhard Rahm \\
Department of Computer Science, University of Leipzig \\
 Interdisciplinary Center for Bioinformatics, University of Leipzig \\ 
\{gross,hartung,thor,rahm\}@informatik.uni-leipzig.de}

\maketitle

\maketitle

\begin{abstract}
Mappings between related ontologies are increasingly used to support data integration and analysis tasks. Changes in the ontologies also require the adaptation of ontology mappings. So far the evolution of ontology mappings has received little attention albeit ontologies change continuously especially in the life sciences. We therefore analyze how mappings between popular life science ontologies evolve for different match algorithms. We also evaluate which semantic ontology changes primarily affect the mappings. We further investigate alternatives to predict or estimate the degree of future mapping changes based on previous ontology and mapping transitions.
\end{abstract}

\setcounter{footnote}{0}

\section{Introduction}
\label{sec:introduction}

Ontologies have become increasingly important in the life sciences~\cite{BS06,L07}. They are used to semantically annotate molecular-biological objects such as proteins or pathways~\cite{TML07}. Different ontologies of the same domain often contain overlapping and related information. For instance, information about mammalian anatomy can be found in NCI Thesaurus~\cite{NCIT} and Adult Mouse Anatomy~\cite{AMA}. Ontology mappings are used to express the semantic relationships between different but related ontologies, e.g., by linking equivalent concepts of two ontologies. 
%Ontology mappings can support other semantic relationship types. For example, an ontology mapping between two sub-ontologies of the Gene Ontology~\cite{GO08} may link \emph{Molecular Functions} that are involved in \emph{Biological Processes}. 

Mappings between related ontologies are useful in many ways, in particular for data integration and enhanced analysis~\cite{NSW+09,JL05}. In particular, such mappings are needed to merge ontologies, e.g., to create an integrated cross-species anatomy ontology such as the Uber ontology~\cite{Uberon}. Anatomy ontology mappings may also be useful to transfer knowledge from different experiments between species \cite{BHR+05}. Furthermore, mappings can help finding objects with similar ontological properties as interesting targets for a comparative analysis. Ontology curators can further find missing ontology annotations and get recommendations for possible ontology enhancements based on mappings to other ontologies. 

Ontologies underly continuous modifications so that new ontology versions are released periodically~\cite{HKR08}. New versions typically incorporate enhanced knowledge, such as additional concepts, relationships, and attribute values. Existing information can also be revised or even deleted. Such ontology changes can invalidate previously determined ontology mappings so that they may have to be re-determined to remain useful. Unfortunately, determining ontology mappings is an expensive process even with the help of semi-automatic ontology matching techniques~\cite{ES07,RB01} that still involve a manual verification of correspondences and a parametrization effort. The importance on determining and adapting ontology mappings is underlined by the popular Ontology Alignment Evaluation Initiative (OAEI)~\cite{OAEI11}. OAEI provides real-world test data sets, in particular for matching the Adult Mouse Anatomy Ontology against the anatomy part of NCI Thesaurus. Unfortunately, the reference mapping of the anatomy task is based on 5 year old ontology versions\footnote{As of 2012, the current reference ontology mapping has been created in 2007.} so that its quality for the current ontology versions remains unclear.

The evolution of ontology mappings has received very little attention so far, especially for the life science domain. For example it is unknown to what degree and how mappings between popular life science ontologies change and how ontology changes affect ontology mappings. There are many ways to compute mappings and it is not clear to what degree different match methods result in differently stable ontology mappings. Finally, we would like to investigate to what degree one can predict future mapping changes based on previously observed ontology and mapping changes. Such information is expected to be useful for deciding about whether a previous ontology mapping is still reliable and up-to-date or whether one has to perform an expensive adaptation of the mapping. 

To address these questions and issues we make the following contributions:

\begin{itemize}
  \item We introduce a generic model for ontology and mapping evolution as well as for their inter-dependencies. The model supports analyzing the impact of ontology evolution on mapping evolution, e.g., what ontology changes lead to the addition or deletion of correspondences in the mapping. (Sec.~\ref{sec:evolution_model}) 
  \item We apply our model to three life science scenarios and evaluate how mappings between popular life science ontologies evolve. We also investigate mapping evolution for different match techniques. (Sec.~\ref{sec:OntoMapEvaluation})
  \item We propose and evaluate two approaches to estimate the number of mapping changes based on previous ontology and mapping changes. (Sec.~\ref{sec:pred})  
\end{itemize}

In Sec.~\ref{sec:prelim} we present preliminaries and outline the general scenario. We describe related work in Sec.~\ref{sec:rel_work} and conclude in Sec.~\ref{sec:summary}.

\section{Preliminaries}
\label{sec:prelim}

\subsection{Ontology, Mapping, and Matching}
\label{sec:base_models}

% Ontology, mappings, ontology versions and mapping versions
%\begin{itemize}
	%\item Ontology: $O = (C,R,A)$, ontology $O$ with concepts $C$, relationships $R$ and attributes $A$
	%\item Mapping: $M_{O1,O2}$, mapping between two ontologies $O1$ / $O2$ consisting of set of correspondences $corr=(c1,c2)$ ($c1 \in O1$, $c2 \in O2$)
	%\item Ontology version: $O_v=(C_v,R_v,A_v,t)$ of version $v$ released at $t$ consisting of concepts $C_v$, relationships $R_v$ and attributes $A_v$; for simplicity we use simple integer numbers 1,2,\ldots to denote the released versions of an ontology
	%\item Mapping version: $M_{O1_u,O2_v}$, mapping between two ontology versions $O1_u$ / $O2_v$, i.e., correspondences refer to valid concepts of $O1_u$ / $O2_v$; for simplicity we assume that only ontologies with the same version number can be matched (see figure later), e.g., $O1_2$ with $O2_2$ results in a mapping version $M_{O1_2,O2_2}$ which we simply reference as $M_{2}$
	%\item for the rest: virtually put contents of the two ontologies (input ontologies) to be matched together and refer to $O$, e.g., union of their concepts; capture ontology changes 'in one place'
%\end{itemize}

In general an {\bf ontology} $O=(C,R,A)$ consists of concepts $C$ which are interrelated by directed relationships $R$. Each concept has an unambiguous identifier such as an accession number. A concept typically has further attributes $a \in A$ to describe the concept, e.g., name, synonyms, or definition. A relationship $r \in R$ forms a directed connection between two concepts and has a specific type, e.g., {\sf is\_a} or {\sf part\_of}. An {\bf ontology mapping} $M_{O1,O2}$ is a set of correspondences $(c1,c2)$ whereby each correspondence interconnects two concepts $c1 \in O1$ and $c2 \in O2$ of the two ontologies. The mapping semantics depends on the intended use case but we assume that all correspondences of a mapping express the same semantic type, e.g., {\sf is-equivalent-to} or {\sf is-related-to}. 

% Ontology mappings are  useful in many ways, in particular for enhanced analysis and annotation of genes, proteins or other objects of interest. For example, such objects may only be assigned to one particular ontology, say GO functions. An ontology mapping between GO functions and GO processes can then be useful to newly assign the objects to the second (process) ontology. Curators could thus use ontology mappings to find missing ontology annotations and get recommendations for possible ontology associations. Conversely, existing ontology associations could be validated against a newly determined ontology mapping in order to locate potential mis-associations reducing data quality. Ontology mappings are also helpful for explorative data analysis, e.g., to find objects with similar ontological properties as interesting targets for a comparative analysis.  
% \textcolor[rgb]{1,0,0}{(brauchen wir $(c1,c2,sim)$ wegen threshold spaeter?)}
% The rapid increase in the number of life science data sources is accompanied by a similar growth in the number of ontologies and mappings between data sources and ontologies.

Since a purely manual creation of ontology mappings is a tedious and labor-intensive task  such mappings are usually determined by semi-automatic {\bf ontology matching} techniques (see Sec.~\ref{sec:rel_work} for Related Work). 
%Matching approaches have been recently studied in diverse scientific and commercial application domains and various match approaches and prototypes (see e.g., \cite{RB01,ES07,R11} for surveys. 
Most matching approaches are metadata-based, i.e., they use the ontology representations themselves to find related concepts, in particular the names of concepts and contextual information like the names of the parent or child concepts within the ontologies. In our evaluation, we will analyze mapping changes for three typical metadata-based matchers (Sec.~\ref{sec:OntoMapEvaluation}).

\subsection{Versioning Scheme}
\label{sec:general_scenario}

\begin{figure}[t]
	\centering
		\includegraphics[width=1.0\textwidth]{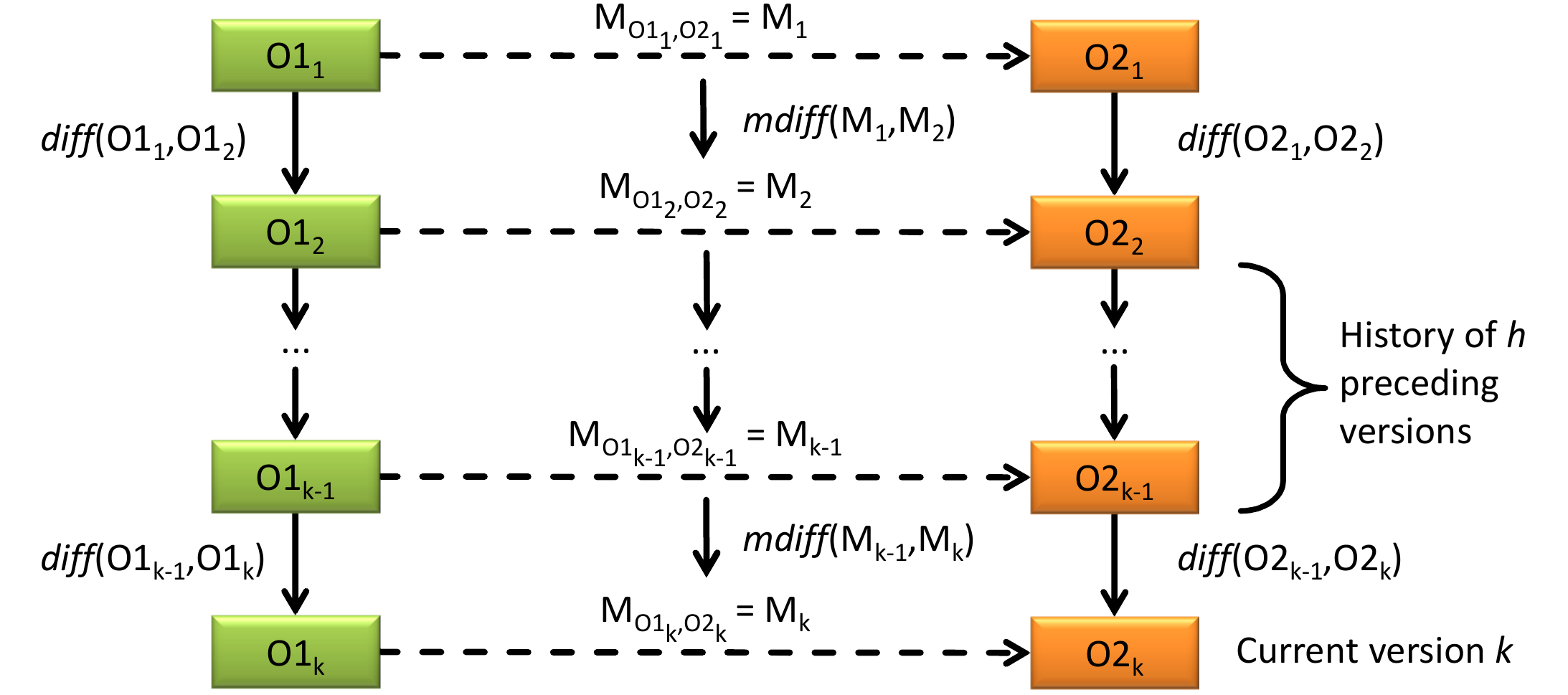}
	\caption{General evolution scheme with multiple ontology and mapping versions}
	\label{fig:general_situation}
\end{figure}

%Ontologies are not static but are frequently evolved to incorporate the newest knowledge of a domain or to adapt to changing application requirements. This is especially true for life science ontologies. For example, the popular Gene Ontology provides monthly releases of its ontologies. 
We  define an {\bf ontology version} $O_v=(C_v,R_v,A_v)$ as a snapshot of an ontology $O$ released at a specific point in time. For simplicity we enumerate the versions with ascending numbers $v=1,2,\ldots$ rather than using the actual release dates. 

Ontology changes affect previously determined ontology mappings so that these mappings should be continuously adapted. Fig.~\ref{fig:general_situation} illustrates the general versioning scheme we adopt in  this paper. There is a series of versions ($v=1 \ldots k$) for a pair of ontologies $O1$ and $O2$ that are connected by an ontology mapping $M_{O1,O2}$. For simplicity we determine ontology mappings only between ontologies of the same version number, i.e., we create mappings $M_v$ only between ontology versions $O1_v$ and $O2_v$ referring to the same specific point in time. 

The difference between two ontology and mapping versions is denoted by $diff(O_v,O_{v+1})$ and $mdiff(M_v,M_{v+1})$, respectively. The next section explains $diff$ and $mdiff$ in more detail.

\section{Change Model for Ontologies and Mappings}
\label{sec:evolution_model}
We first describe our change model for ontologies and mappings and categorize the changes into different groups. We also propose simple change ratio indicators to  assess the evolution intensity between successive ontology and mapping versions. We then propose indicators to assess the impact of ontology changes on ontology mappings.  

\subsection{Ontology Changes}
\label{sec:onto_evol}

\begin{table}[t]
\begin{center}

\begin{tabular}{|l|l|}
\hline
Change operation & Type\\ 
\hline \hline 
Insertion of a new concept to $O_{v+1}$ & \multirow{5}{*}{Information extension}\\
Insertion of a subgraph to a concept & \\
Insertion of new relationship in $O_{v+1}$ & \\
Addition of an attribute (to an existing concept) & \\
Mark concept as non-obsolete & \\
\hline
Deletion of a concept in $O_{v}$ & \multirow{5}{*}{Information reduction}\\
Removal of a subgraph  & \\
Deletion of an relationship in $O_{v}$ & \\
Deletion of an existing attribute & \\ 
Mark concept as obsolete & \\
\hline
Split concept of $O_{v}$ into multiple concepts in $O_{v+1}$ & \multirow{5}{*}{Information revision} \\
Merge concepts of $O_{v}$ into a single concept in $O_{v+1}$& \\
Concept substitution  & \\
Move concept &  \\
Change attribute value & \\
\hline 
\multicolumn{2}{c}{ } 
\end{tabular}
\caption{COntoDiff change operations (including their categorization in three groups) for ontology evolution $O_{v} \mapsto O_{v+1}$.}
\label{tab:changes}
\end{center}
\end{table}

\begin{figure}[t]
	\centering
		\includegraphics[width=1\textwidth]{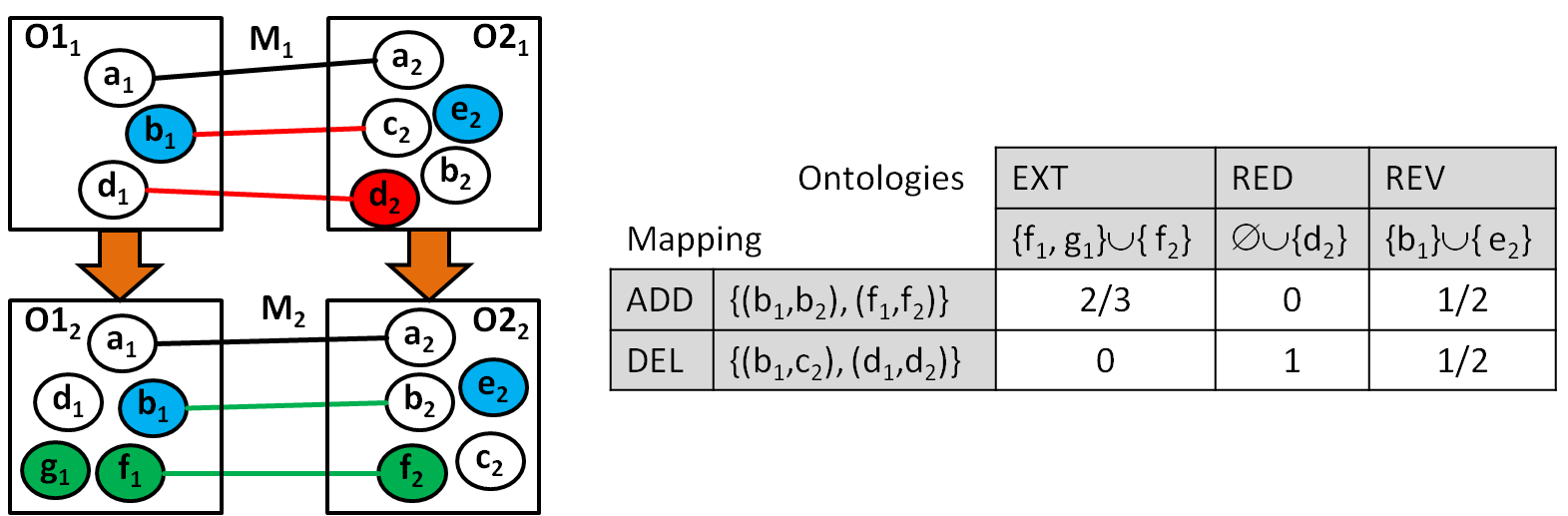}
	\caption{{\bf left:} Example evolution of two ontologies and a mapping. Concepts $b_1$ and $e_2$ have been revised, $d_2 \in O2$ has been removed, and $g_1$, $f_1$, and $f_2$ have been added during the evolution from version $v=1 \mapsto 2$. The mapping change between $O1$ and $O2$ comprises two new correspondences ($(b_1,b_2)$, $(f_1,f_2)$) and two removed correspondences ($(b_1,c_1)$, $(d_1,d_2)$). {\bf right:} Impact matrix of ontology and mapping changes.}
	\label{fig:general_situation_example}
\end{figure}
We start by defining what changes can occur between successive ontology versions $O_{v}$ and $O_{v+1}$. Our model is based on the COntoDiff algorithm described in~\cite{HGR10}. COntoDiff computes the difference $diff(O_{v},O_{v+1})$ between an old and a new version of an ontology and consists of the set of change operations that -- when applied to $O_{v}$ -- transform the old into the new version. Basic change operations are concept and attribute additions or deletions. COntoDiff also determines more complex changes such as merging or splitting of concepts or the addition/deletion of subgraphs. 
 
Table~\ref{tab:changes} lists all considered change operations and additionally categorizes them into one of three groups. The first group contains information extending operations that add information in $O_{v}$ such as new concepts, relationships or attribute values. The second group, information reduction, includes change operations that remove information from $O_{v}$. All other operations including split and merge changes belong to the revise group.
\\
     
% \marking{Begriffe information increasing/reducing/revising noch verbesserungsfaehig \ldots}
% following changes for ontologies: \emph{addLeaf}, \emph{addInner}, \emph{addSubGraph}, \emph{delLeaf}, \emph{delInner}, \emph{delSubGraph}, \emph{merge}, \emph{split}, \emph{toObsolete}, \emph{move}, \emph{addA}, \emph{delA}, \emph{chgAttValue}, \emph{addR} and \emph{delR}. In our general scenario displayed in Fig.~\ref{fig:general_situation} we determine evolution mappings between succeeding ontology versions, i.e., between $O1_i$-$O1_{i+1}$ ($diff(O1_{i},O1_{i+1})$) and $O2_i$-$O2_{i+1}$ ($diff(O2_{i},O2_{i+1})$), respectively.
For a quantitative change analysis we assign concepts both from $O_{v}$ and $O_{v+1}$ based on their change operations to one of the following sets:

\begin{itemize}
\item {\bf Extension set:} $Ext(O_{v \mapsto v+1})$ = set of concepts in $O_{v} \cup O_{v+1}$ where all concept-related change operations are information extending.  
\item {\bf Reduction set:} $Red(O_{v \mapsto v+1})$ = set of concepts in $O_{v} \cup O_{v+1}$ where all concept-related change operations are information reducing.
\item {\bf Revision set:} $Rev(O_{v \mapsto v+1})$ =  set of concepts in $O_{v} \cup O_{v+1}$ that are involved in at least one change operation but belong neither to $Ext$ nor to $Red$. Each concept is thus related to a revise operation or is related to both extending and reducing operations. 
\end{itemize}

All other concepts remain unchanged, i.e., they are not affected by any change operation. Fig.~\ref{fig:general_situation_example} illustrates an evolution example for two ontologies $O1$ and $O2$. For example, the evolution from $O2_1$ to $O2_2$ might contain three change operations: insertion of concept $f_2$, deletion of concept $d_2$, and an attribute value change for concept $e_2$. The three concepts are thus assigned to $Ext$, $Red$, and $Rev$, respectively, i.e., $Ext(O2_{1 \mapsto 2})=\{f_2\}$, $Red(O2_{1 \mapsto 2})=\{d_2\}$, and $Rev(O2_{1 \mapsto 2})=\{e_2\}$. All other concepts of Fig.~\ref{fig:general_situation_example} are not affected by the change operations.

% \marking{Hier ist etwas unschoen, dass das Beispiel so simpel ist, dass es nur eine change operation pro concept gibt}  
% \marking{Vielleicht in dem Beispiel auch mit $O1_{old}$ und $O1_{new}$ -- das gleiche fuer O2 -- arbeiten?}
% To study the influences we first group determined changes into the following change groups (classes):  
% \begin{itemize}	
% 	\item additions in input ontologies between versions $i$ and $j$: $Add_{O,i,j}$ contains all ontology concepts involved in an information extending change
% 	\item deletions in input ontologies between versions $i$ and $j$: $Del_{O,i,j}$ contains all ontology concepts involved in an information reducing change
% 	\item revisions in input ontologies between versions $i$ and $j$: $Rev_{O,i,j}$ contains all ontology concepts involved in a revise change
% 	\item stable (unchanged) part in input ontologies between versions $i$ and $j$: $Unc_{O,i,j}$ contains all unchanged ontology concepts
% \end{itemize}

The size of the three concept sets $Ext$, $Red$, and $Rev$ quantitatively characterizes the degree of change during the evolution from $O_{v}$ to $O_{v+1}$. We therefore define the {\bf ontology change ratio} as follows:
$$ OCR(O_{v \mapsto v+1}) = \frac{|Ext(O_{v \mapsto v+1}) \cup Red(O_{v \mapsto v+1}) \cup Rev(O_{v \mapsto v+1})| }{ |O_{v} \cup O_{v+1}| }$$

The ontology change ratio for $O2$ of our running example (Fig.~\ref{fig:general_situation_example}) is thus $OCR(O2_{1 \mapsto 2}) = |\{f_2, d_2, e_2\}|/|\{a_2,b_2,c_2,d_2,e_2,f_2\}|=0.5$.
% \marking{overall change factor for input ontologies: $\frac{|Add_{O,i,j}|+|Del_{O,i,j}|+|Rev_{O,i,j}|}{|O_i \cup O_j|}$ (change factors only for $Add$, $Del$, $Rev$ possible)}

\subsection{Mapping Changes}
\label{sec:map_evol}
For ontology mapping evolution we employ a simple model that distinguishes between the addition and deletion of correspondences. Thus, between two consecutive mapping versions $M_{v}$ and $M_{v+1}$ we consider whether a new correspondence has been added ($Add$) or a previous one has been removed ($Del$). 
%Our model thereby ignores any confidence values (as computed by the ontology matching algorithm) but considers selected match correspondences only. We leave the analysis of evolutionary changes of confidence values as subject for future work. 
We group changed correspondences into the following sets:

\begin{itemize}
\item {\bf Addition set:} $Add(M_{v \mapsto v+1}) = M_{v+1} \backslash M_{v}$
\item {\bf Deletion set:} $Del(M_{v \mapsto v+1}) = M_{v} \backslash M_{v+1}$ 
\end{itemize}

All other correspondences appear in both mapping versions and are thus unchanged. Based on the introduced sets we define the {\bf mapping change ratio} as follows: 
$$ MCR(M_{v \mapsto v+1}) = \frac{|Add(M_{v \mapsto v+1}) \cup Del(M_{v \mapsto v+1})|}{|M_{v} \cup M_{v+1}|}$$

In the example of Fig.~\ref{fig:general_situation_example} there are two new correspondences, i.e., $Add(M_{1 \mapsto 2})=\{(b_1,b_2),(f_1,f_2)\}$. and two deleted correspondences, $(b_1,c_2)$ and $(d_1,d_2)$. Since there is one unchanged correspondence $(a_1,a_2)$, the mapping change ratio\\$MCR(M_{1 \mapsto 2})$ equals $4/5$.
% Analogous to ontologies, we determine changes between succeeding mapping versions (e.g., between $M_{1}$ and $M_{2}$ in Fig.~\ref{fig:general_situation}). 
% Note that we do not determine the changes between mapping version $M_{k-1}$ and $M_{k}$ since we do not know the content of the last mapping version. It will be task our estimation (see following Section) to a priori estimate how many changes will occur between both versions.
% \begin{itemize}	
% 	\item additions between two mapping versions $i$ and $j$: $Add_{M,i,j}$ contains all added correspondences
% 	\item deletions between two mapping versions $i$ and $j$: $Del_{M,i,j}$ contains all deleted correspondences
% 	\item stable (unchanged) part between two mapping versions $i$ and $j$: $Unc_{M,i,j}$ contains all unchanged correspondences
% 	\end{itemize}
% \subsection{Change Model}
% \label{sec:change_model}
% 
% \subsection{Classification of Changes}
% \label{sec:change_classification}
 
\subsection{Impact of Ontology on Mapping Changes}

\label{sec:ontMapChangeAssociations}

To determine how ontology changes influence or trigger mapping changes it is useful to interrelate the different kinds of ontology changes and mapping changes.  For this purpose, we interrelate the three sets of changed concepts ($Ext$, $Red$, $Rev$) with the two sets of changed correspondences ($Add$, $Del$). We will define six corresponding indicators and use them  for both analyzing mapping evolution (see Sec.~\ref{sec:OntoMapEvaluation}) as well as for predicting mapping changes for new ontology versions (see Sec.~\ref{sec:pred}). 

The {\bf impact ratio} is the share of changed concepts that actually had an impact on the correspondences. For any set of ontology changes $O_{Ch}$ ($Ext$, $Red$, or $Rev$) and mapping changes $M_{Ch}$ ($Add$ or $Del$) it is defined as follows:
$$ IR (O_{Ch},M_{Ch}) = \frac{|\{ c \in  O_{Ch} | \exists c': (c,c') \in M_{Ch} \vee (c',c) \in M_{Ch} \}|}{|O_{Ch}|} $$

% We define the {\bf impact association set} for an ontology change set $O_{Ch}$ ($Ext$, $Red$, or $Rev$) and a mapping change set $M_{Ch}$ ($Add$ or $Del$) as follows:
% $$ Imp (O_{Ch},M_{Ch}) =  $$
% $$ Imp (Ext(O1_{v \mapsto v+1}), Ext(O2_{v \mapsto v+1}), Add(M_{v \mapsto v+1}))  = \{ c \in   Ext(O1_{v \mapsto v+1}) \cup Ext(O2_{v \mapsto v+1}) | \exists c': (c,c') \in Add(M_{v \mapsto v+1}) \wedge (c',c) \in Add(M_{v \mapsto v+1}) \} $$

For example, to determine which fraction of additive ontology changes led to new correspondences we determine the impact ratio for $O_{Ch} = Ext(O1_{1 \mapsto 2}) \cup Ext(O2_{1 \mapsto 2})$ and $M_{Ch} = Add(M_{1 \mapsto 2})$. For the example in Fig.~\ref{fig:general_situation_example},  two ($f_1$ and $f_2$) out of the three $Ext$-concepts appear in the set of added correspondences, i.e., the changes in these two concepts had an impact on the mapping. Therefore $IR(Ext,Add)$ equals $\frac{2}{3}$.

% The resulting impact association set is therefore $Imp (Ext,Add) = \{ f_1, f_2 \}$. 

One would expect that $Ext$ concepts mostly lead to correspondence additions whereas $Red$ concepts usually account for correspondence deletions. However, as we will see in our evaluation (see Sec.~\ref{sec:OntoMapEvaluation}), $Ext$ concepts may also trigger correspondence deletions and $Red$ concepts may lead to new correspondences depending on the match technique.

\section{Analysis of Mapping Evolution}
\label{sec:OntoMapEvaluation}
After introducing the experimental setup, we analyze ontology and mapping evolution for different life science scenarios. We then compare mapping evolution for different match strategies and evaluate the impact of ontology changes on mapping changes. 

\subsection{Setup}
We consider three mapping scenarios: 
\begin{itemize}
	\item \textit{Anatomy}: map Adult Mouse Anatomy Ontology (MA) to the anatomy part of NCI Thesaurus (NCITa)
	\item \textit{Molecular~Biology}: map the two Gene Ontology\cite{GO08} sub-ontologies Molecular Functions (MF) and Biological Processes (BP) 
	\item \textit{Chemistry}: map Chemical Entities of Biological Interest (ChEBI)~\cite{Chebi} to NCI Thesaurus (NCIT)
\end{itemize}
For each input ontology we map 10 versions on a half year basis between 2006-06 and 2010-12 with each other. 
%If there was no version from June or December available, we used the last valid version. We produced mappings by matching always corresponding ontology versions from the same point in time. 
We use the following meta-data based matchers to compute the confidence (similarity) for any concept pair of two ontologies:
 
\begin{itemize}
  \item \textit{Name}: String (trigram) similarity of concept names
  \item \textit{NameSyn}: Maximal string (trigram) similarity of names and synonyms 
  \item \textit{Context}: String (trigram) similarity of the concatenated parent, concept, and children names
\end{itemize}

In this study we focus on the evolution of ontology mappings and do not evaluate the quality of matching. The choice of match strategies is based on previous studies where matching on  concept names and synonyms achieved high quality especially for anatomy ontologies ~\cite{GNM09,GHKR11}. To obtain precise results we need to select the most likely correspondences exceeding a certain confidence threshold. We applied a default confidence threshold of \textit{0.6}; for the \textit{NameSyn} matcher, we also considered a stricter threshold of \textit{0.8}. Moreover, for each input ontology concept, we only select the top correspondences in a small delta range (MaxDelta selection~\cite{DR02}). 

\subsection{Ontology and Mapping Evolution}

\begin{figure}[t]
	\centering
		\includegraphics[width=1.0\textwidth]{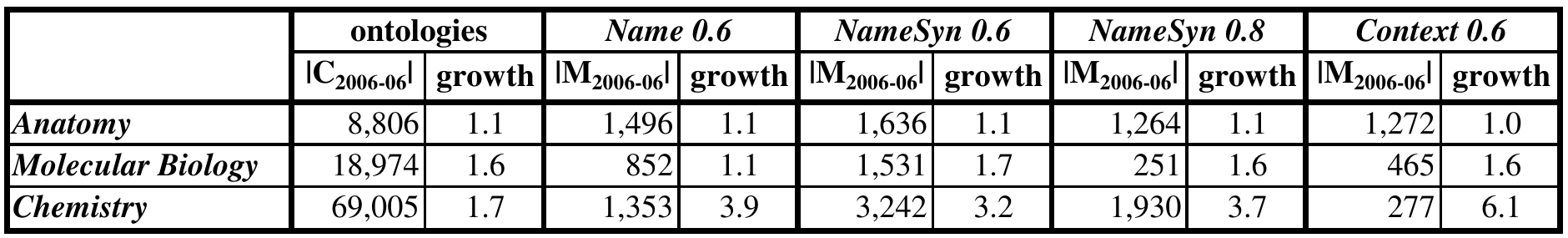}
	\caption{Ontology and mapping growth factors. Number of concepts ($|C_{2006-06}|$) and number of mapping correspondences ($|M_{2006-06}|$) in the first considered version. $|C|$ is the sum of domain and range ontology size for each match problem. Growth factors compare the first (2006-06) and last (2010-12) considered version.}
	\label{fig:onto_mapp_growth}
\end{figure}

Fig.~\ref{fig:onto_mapp_growth} gives an overview about the ontology and mapping sizes as well as their growth between June 2006  and Dec. 2010. For \textit{Anatomy}, the combined size of concepts in domain and range ontology ($|C|$) grew only slightly by a factor 1.1 to almost 10,000 concepts. By contrast, $|C|$ increased by 60 - 70 \% to ~30,000 and ~120,000 concepts for \textit{Molecular~Biology} and \textit{Chemistry}. In two of the three scenarios (\textit{Anatomy} and \textit{Molecular~Biology}), the mappings grow similarly strong as the ontologies while the \textit{Chemistry} mappings grew  by up to a factor 6. %making a detailed analysis necessary.
The especially high mapping growth  for the \textit{Context} matcher seems influenced by its very small mapping size which in turn is caused by its need to find similar names not only for the concepts but also for their parent and child concepts. Comparing the results for \textit{NameSyn} with two different thresholds, we find that a higher threshold produces smaller mappings and achieves only a relatively small coverage, especially for \textit{Molecular~Biology}. For \textit{Molecular~Biology}, the \textit{Name} matcher proved to determine the most stable mappings. 
%This might be interesting concerning the impact of ontology evolution. For instance, if ontology parts change where no correspondences are located, it might not result in a mapping change. By contrast, mappings covering strongly evolving ontology parts might be heavily changed over time.

\begin{figure}[b]
	\centering
		\includegraphics[width=1.0\textwidth]{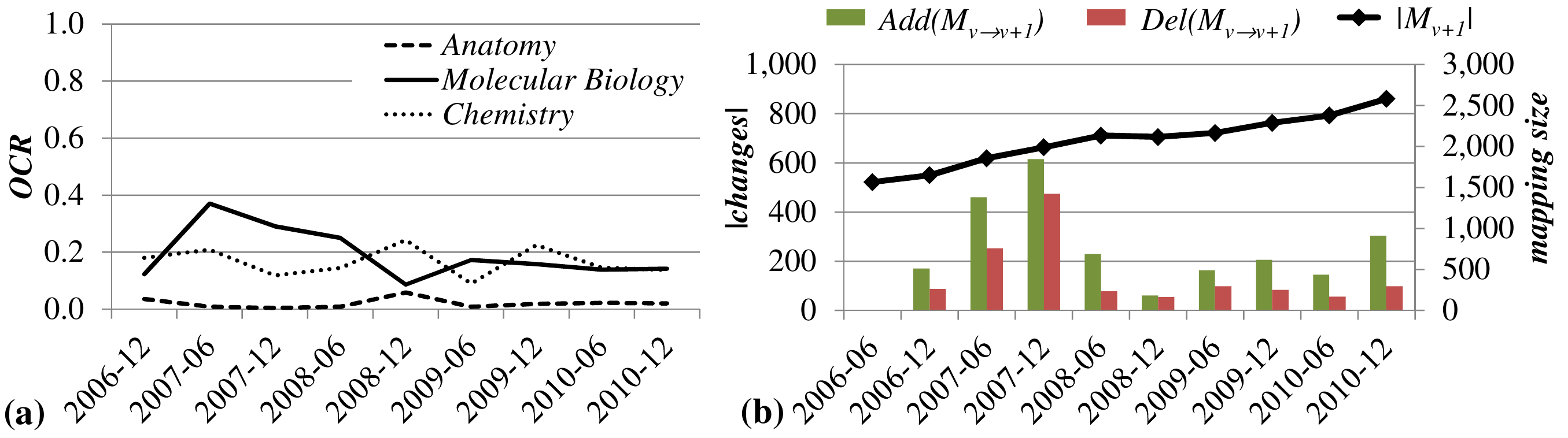}
	\caption{(a) Ontology change factors. (b) Mapping evolution for \textit{NameSyn 0.6} matcher in \textit{Molecular~Biology} example.}
	\label{fig:onto_mapp_evolution}
\end{figure}

Fig.~\ref{fig:onto_mapp_evolution}(a) shows ontology change factors (see Sec.\ref{sec:ontMapChangeAssociations}) between succeeding versions for the three domains during the 5-year observation period. For \textit{Anatomy} there were only few changes over time compared to the other two domains. \textit{Molecular~Biology} shows high change rates until 2007 (nearly 40\%). From 2008 on, change rates are comparable to those of \textit{Chemistry} (around 20\%). Fig.~\ref{fig:onto_mapp_evolution}(b) illustrates more detailed mapping evolution results for \textit{NameSyn 0.6} in \textit{Molecular~Biology}. In general, correspondence additions dominate leading to a final mapping size of more than 2,500 correspondences. But there has also been a considerable number of deletions. In 2007-12 nearly 500 correspondences were removed from the mapping. This shows that there can be very heavy mapping changes.

\subsection{Comparison of Match Strategies}

\begin{figure}[t]
	\centering
		\includegraphics[width=1.0\textwidth]{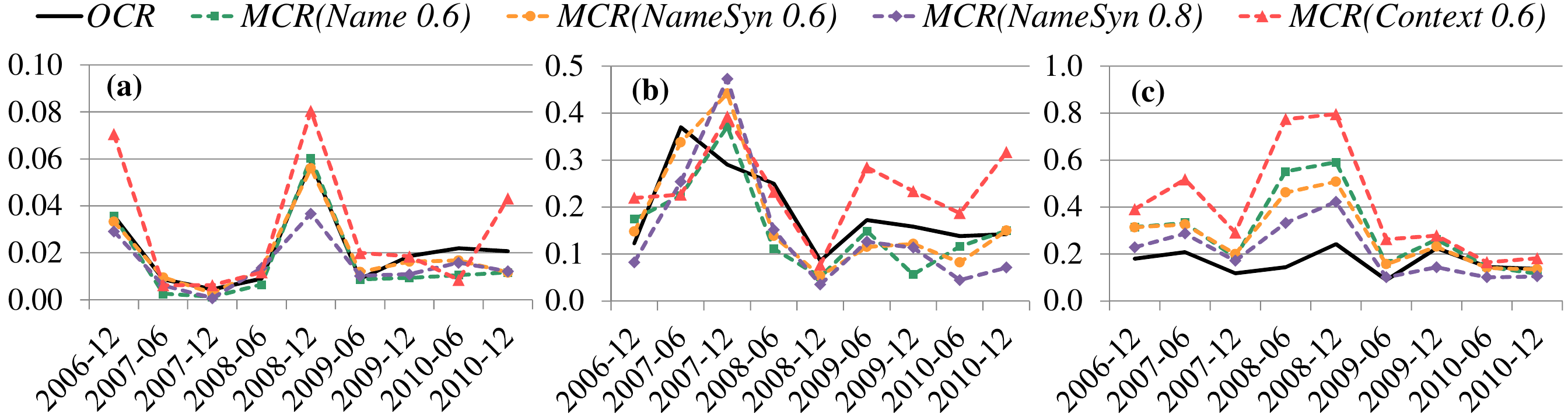}
	\caption{Ontology and mapping change factors for three life science domain examples (a) \textit{Anatomy}, (b) \textit{Molecular~Biology}, (c) \textit{Chemistry}}
	\label{fig:change_factors_allDomains}
\end{figure}

\begin{figure}[b]
	\centering
		\includegraphics[width=1.0\textwidth]{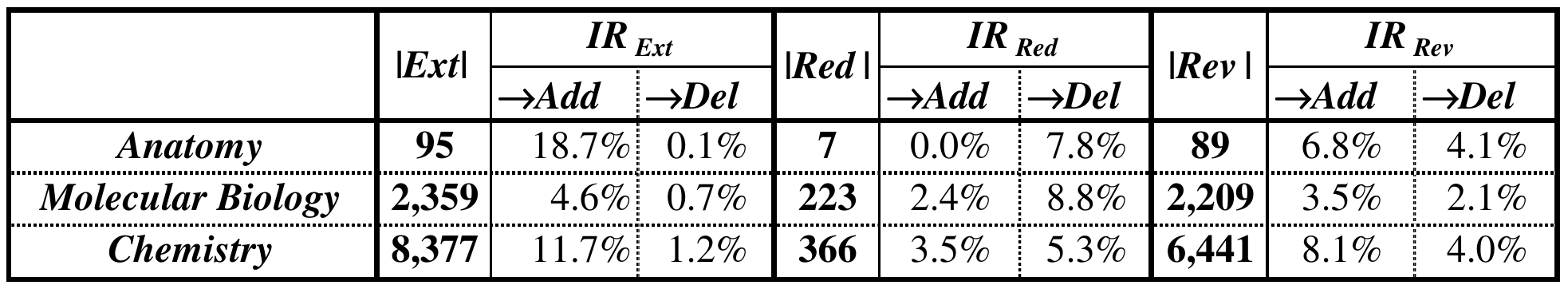}
	\caption{Impact of ontology concept changes ($Ext$, $Red$, $Rev$) on mapping changes ($Add$, $Del$) for \textit{NameSyn 0.6}. Average values for absolute change number ($\left|Ext\right|$, $\left|Red\right|$, $\left|Rev\right|$) and impact association ratios ($IR(O_{Ch},M_{Ch})$ displayed as percentage) over all considered versions}
	\label{fig:evolutionImpact_for_nameSyn}
\end{figure}

To analyze the mapping stability for different match strategies in more detail, we examine a possible correlation between ontology and mapping changes over time. We therefore compute ontology and mapping change factors for all three match scenarios and the four match strategies (Fig.~\ref{fig:change_factors_allDomains}~a-c). For \textit{Anatomy}, ontologies and mappings only slightly changed (see y-axis range), while the other two scenarios experience  a surprisingly high degree of mapping changes between 10 and  80 \%.  Except for \textit{Chemistry} we observe  a strong correlation between the ontology change factor (black continuous line) and the mapping change factors of the different match strategies(colored dashed lines).  The \textit{Name} matcher was relatively stable in general while the \textit{Context} matcher was most heavily influenced by ontology evolution. This especially holds for \textit{Chemistry} where 80\% of the \textit{Context} mappings changed in 2008. The reason for the relative instability of \textit{Context} is mainly in its use of more ontological information that can change, i.e., changes on both parent and child concepts have an influence. For instance, moving a concept from one parent concept to another might completely change a concept's context. For \textit{Molecular~Biology} the mappings, (especially \textit{NameSyn}), changed heavily in 2007-12, although the maximum ontology evolution already occurred in 2007-06. This results from successive modification of GO-BP and GO-MF in 2007. The combined changes in both sub-ontologies seem to have led to numerous mapping changes in 2007-12. 

\subsection{Impact of Ontology on Mapping Changes}
Fig.~\ref{fig:evolutionImpact_for_nameSyn} illustrates the real impact of ontology changes (\emph{Ext}, \emph{Red}, \emph{Rev}) on mapping changes (\emph{Add}, \emph{Del}). We exemplarily show results for \textit{NameSyn 0.6} and computed the average over all versions. The table shows the number of changed concepts as well as the ratio having impact on mapping changes ($IR$). First, we can observe that a high number of ontology extensions, reductions and revisions has no impact on the ontology mappings (\textgreater80\%). This is due to a limited match coverage since changed ontology parts that are not covered by the ontology mapping do not result in mapping changes. Second, extending ontology changes (\emph{Ext}) primarily cause correspondence additions and no or only few correspondence deletions for all three scenarios. Third, \emph{Red} concepts are primarily involved in correspondence deletions but also in some additions. The latter might result from specific matcher characteristics. Imagine a concept loses a synonym and also the correspondence based on this synonym. This can enable a new correspondence by relating the concept to another one than before. Thus, a synonym deletion can lead to a correspondence deletion and addition in one evolution step. Finally, revised concepts (\emph{Rev}) trigger both, \emph{Add} and \emph{Del}. This is intuitive since revised concepts might have been extended and reduced in one evolution step (e.g., attribute addition and deletion). In general, ontology revisions account for a high share of mapping changes while deletions play only a minor role. 

\subsection{Summary}

We evaluated ontology and mapping evolution for three real-world life science domains (\textit{Anatomy}, \textit{Molecular~Biology} and \textit{Chemistry}) and took four match-strategies into account. The analysis results show that especially \textit{Molecular~Biology} and \textit{Chemistry} underlie heavy ontology extensions and revisions whereas \textit{Anatomy} is relatively stable. Since existing knowledge is mainly extended or revised, we find only few ontology reducing changes for all domains. 
Ontology evolution heavily influenced mappings computed by different metadata-based match strategies. Especially, the structural matcher \textit{Context} produced rather unstable results whereas mappings based on the \textit{Name} matcher are relatively stable. As expected, ontology extensions primarily lead to correspondence additions and information reducing ontology changes primarily lead to the removal of correspondences. Ontology revisions play an important role and result in both the addition and deletion of correspondences.    

%\textcolor[rgb]{0.5,0.5,0.5}{
%\begin{itemize}
%\item Show change intensity in input ontologies as well as in resulting mappings
%\item Compare change factors to draw conclusions about the dependency between changes in ontologies and mappings
%\item How strong are changes propagated into the mappings? Differences between matchers and match problems.
%\item Results: existence of trends which one may use for estimating future changes, differences between domains: anatomy stable in comparison to the molecular biology and chemistry, good reuse potential for anatomy match case (few changes large parts remain stable), two other domains: still a lot of knowledge additions/revisions which lead to mapping changes
%\end{itemize}}

\section{Mapping Change Estimation}
\begin{table}[t]
	\centering
		\includegraphics[width=0.9\textwidth]{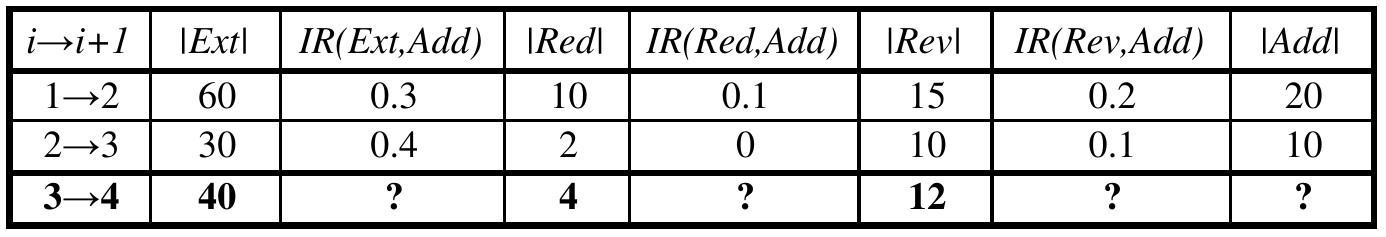}
	\caption{Example prediction scenario.}
	\label{tab:estimation_example_values}
\end{table}
\label{sec:pred}
We now present two methods to estimate the number of changes in a new mapping version. By predicting future mapping changes we can give recommendations to users if it might be necessary to recompute their mappings. This seems especially useful when one must decide about performing an expensive manual mapping adaption or not. We first describe the methods and then comparatively evaluate their quality on our mapping problems.

\subsection{Prediction Methods}
\label{sec:prediction_models}
The general task of estimating mapping changes is the following. After the release of two new ontology versions $O1_k$/$O2_k$ we like to predict the number of mapping changes ($|Add(M_{k\text{-}1 \mapsto k}|$,$|Del(M_{k\text{-}1 \mapsto k}|$) which will occur between the mapping versions $M_{k\text{-}1}$ and $M_k$, i.e., we like to know how strong  mapping $M$ is likely to change due to modifications in $O1$/$O2$. For this estimation we can access the content of the previous $h$ ontology/mapping versions (\emph{v=k-h,\ldots,k-1}) and their diff results. In the following we describe two prediction methods, namely \emph{Mapping-based Estimation (ME)} and \emph{Impact-based Estimation (IE)}. The synthetic example in Table~\ref{tab:estimation_example_values} will be used for illustration.

\emph{Mapping-based Estimation} In this approach the prediction only uses information about previous mapping changes but not about the underlying ontology changes. The estimation for $|Add(M_{k\text{-}1 \mapsto k})|$ and $|Del(M_{k\text{-}1 \mapsto k})|$ is the weighted average of the number of changes observed in the last $h\text{-}1$ version changes of the mapping. We can use different functions $w$ to weight the version changes:
\begin{center}
$|Add(M_{k\text{-}1 \mapsto k})| = \sum_{v=k-h+1}^{k-1}{w_i\cdot|Add(M_{v\text{-}1 \mapsto v})|}$\\
$|Del(M_{k\text{-}1 \mapsto k})| = \sum_{v=k-h+1}^{k-1}{w_i\cdot|Del(M_{v\text{-}1 \mapsto v})|}$
\end{center}
For our example in Table~\ref{tab:estimation_example_values} we like to make a prediction for the number of added correspondences between version 3 and 4 ($|Add(M_{3 \mapsto 4})|$) using the versions 1--3 (\emph{h=3}). We use a quadratic weighting function with the following weights for the two previous version changes: $\frac{1}{5}$ and $\frac{4}{5}$ . We would thus estimate $|Add(M_{3 \mapsto 4})|$ = $\frac{1}{5}\cdot 20+\frac{4}{5}\cdot 10 = 12$ with the \emph{ME} method.

\begin{figure}[t]
  \centering
  \subfigure[]{
    \includegraphics[width=0.4\textwidth]{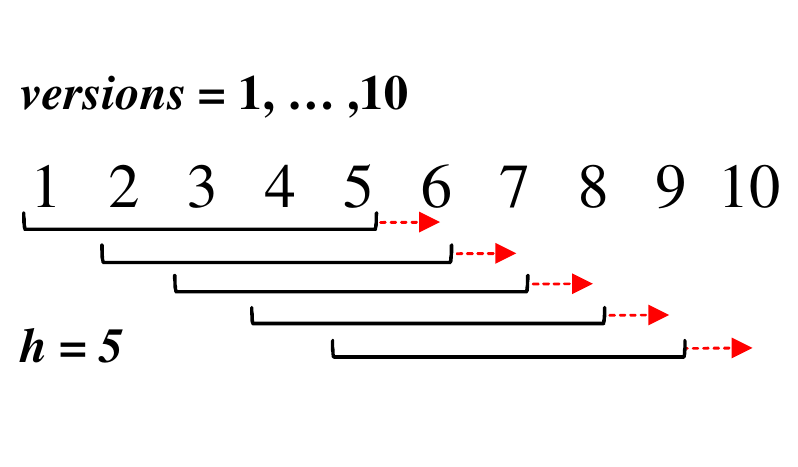} 
  }
  \subfigure[]{
    \includegraphics[width=0.5\textwidth]{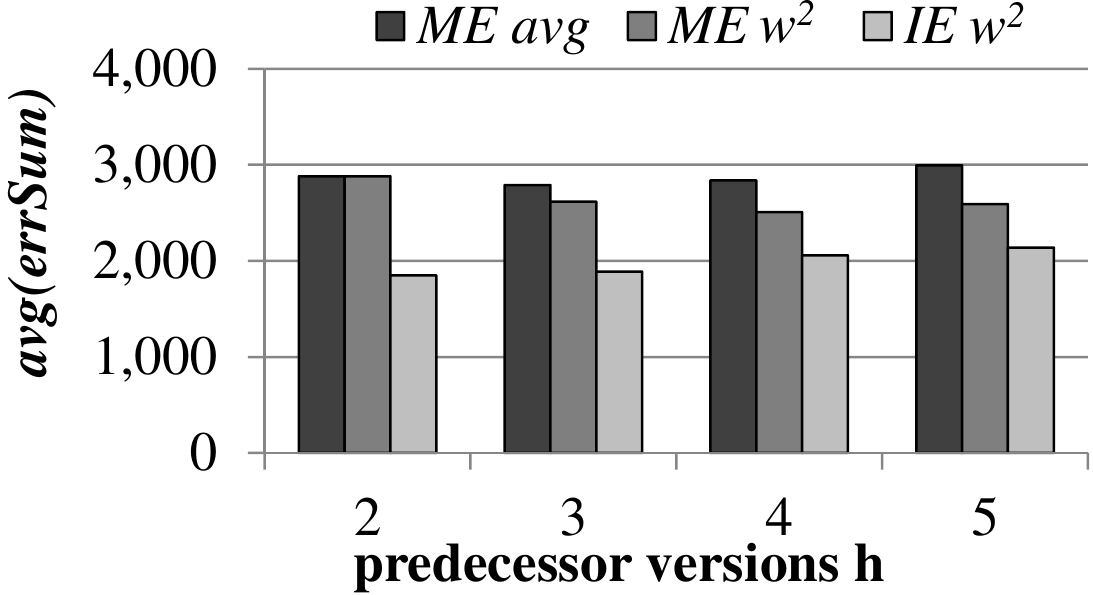} 
  }
  \caption{Prediction analysis (a) Example for predicting the successor version (red dotted line) on the basis of a window of 5 predecessor versions (h=5), (b) Average error sum ($avg(errSum))$ of false predictions for $h=2\ldots5$ for three methods $ME avg$, $ME w\textsuperscript{2}$, $IE w\textsuperscript{2}$}
  \label{fig:slidingWindow}
\end{figure}

\emph{Impact-based Estimation} The idea behind impact-based estimation is to use knowledge about the impact of ontology on mapping changes to estimate the number of correspondence changes. We assume that the number of added/deleted correspondences can be expressed as a linear combination of the observed ontology changes having an impact:
\begin{eqnarray*}
	|Add(M_{k\text{-}1 \mapsto k})| &=& \beta\cdot(\texttt{agg}(IR(Ext,Add))\cdot|Ext(O_{k\text{-}1 \mapsto k})|\\&+&\texttt{agg}(IR(Red,Add))\cdot|Red(O_{k\text{-}1 \mapsto k})|\\&+&\texttt{agg}(IR(Rev,Add))\cdot|Rev(O_{k\text{-}1 \mapsto k})|)
\end{eqnarray*}
\begin{eqnarray*}
	|Del(M_{k\text{-}1 \mapsto k})| &=& \beta\cdot(\texttt{agg}(IR(Ext,Del))\cdot|Ext(O_{k\text{-}1 \mapsto k})|\\&+&\texttt{agg}(IR(Red,Del))\cdot|Red(O_{k\text{-}1 \mapsto k})|\\&+&\texttt{agg}(IR(Rev,Del))\cdot|Rev(O_{k\text{-}1 \mapsto k})|)
\end{eqnarray*}
For both formulas we need two specify two parameters. First, we need to determine the impact ratios ($IR$) which indicate how strong ontology changes will influence the mapping in the current version change. Since we consider the last $h\text{-}1$ version changes, we need to aggregate the observed impact ratios into a common value (\texttt{agg} function), e.g., by a normal or weighted average. Second, we need to determine the $\beta$ parameter which performs an error correction on the result. In particular, for each version change we calculate the estimated value using the linear combination formula with the impact ratios observed. We then compare the estimation with the correct result and compute an error ratio between both. We finally take the average of all computed error ratios as our $\beta$. 

For our example we need to determine three impact ratios. We will use the same quadratic weighting as for \emph{ME} to compute a weighted average. Thus, for $IR(Ext,Add)$ we would determine a value of $\frac{1}{5}\cdot 0.3+\frac{4}{5}\cdot 0.4=0.38$ ($IR(Red,Add)=0.02$ and $IR(Rev,Add)=0.12$). We further calculate the error ratio for each version change, e.g., for $1\mapsto2$ the estimated result is $0.3\cdot 60+0.1\cdot 10+0.2\cdot 15=22$. A comparison with the correct number of correspondence additions ($|Add(M_{1 \mapsto 2})|=20$) results in a ratio of $\frac{20}{22} \approx 0.91$ ($2\mapsto 3: 0.77$). Thus, the average error ratio over all version changes is $\beta=0.84$ resulting in an estimation of $|Add(M_{3 \mapsto 4})|$ = $0.84\cdot(0.38\cdot 40+0.02\cdot 4+0.12\cdot 12) \approx 14$.

\subsection{Evaluation}
We now apply our two estimation methods to predict how many correspondence additions and deletions might occur in a future mapping. We use the same datasets as before (see Sec. \ref{sec:OntoMapEvaluation}). For the map-based method (\emph{ME}), we applied two different weight functions: average (\emph{avg}) and quadratic weighting  average (\emph{w\textsuperscript{2}}). For the Impact-based method (\emph{IE}), we only show results for $w\textsuperscript{2}$ since this showed to be more effective. To get an overview how accurate both methods are and how many versions are required for good estimation, we performed the following experiment.

We predict the last five mapping versions using several numbers of predecessor versions ($h=2\ldots5$). Fig.~\ref{fig:slidingWindow}(a) exemplarily shows the experimental scenario for $h=5$. We produce five results per $h$ for the \emph{ME avg}, \emph{ME w\textsuperscript{2}} and \emph{IE w\textsuperscript{2}} prediction methods. For each $h$ and method we compute an error sum (\emph{errSum}: sum of absolute differences between correct (\emph{CR}) and predicted result (\emph{PR})) over all prediction results for three matchers (\textit{Name 0.6}, \textit{NameSyn 0.6}, \textit{Context 0.6}) and all match scenarios. To better compare the methods and to study the influence of $h$ we compute average error sums which are displayed in Fig.~\ref{fig:slidingWindow}(b). For $h=2$, \emph{ME avg} and \emph{ME w\textsuperscript{2}} produce the same results since they only consider one mapping version diff. For a higher number of predecessor versions ($h>2$) \emph{ME w\textsuperscript{2}} produces smaller errors. Overall \emph{IE w\textsuperscript{2}} is more effective than both \emph{ME} methods, i.e., using information about ontology evolution as well, seems to be more informative and thus leads to more accurate results. Especially, only considering the recent past (small $h$) suffices to make a good estimation with our impact-based method \emph{IE}.

%\textcolor[rgb]{1,0,0}{EVENTUELL NOCH: In average 3000 errors for all mappings (3 scenarios * 3 matchers). Average sum of all mapping sizes? Average sum of all correct addCorr and delCorrs? fuer Fig7b doch besser einen normalisierten wert?)}
\begin{figure}[t]
	\centering
		\includegraphics[width=1.0\textwidth]{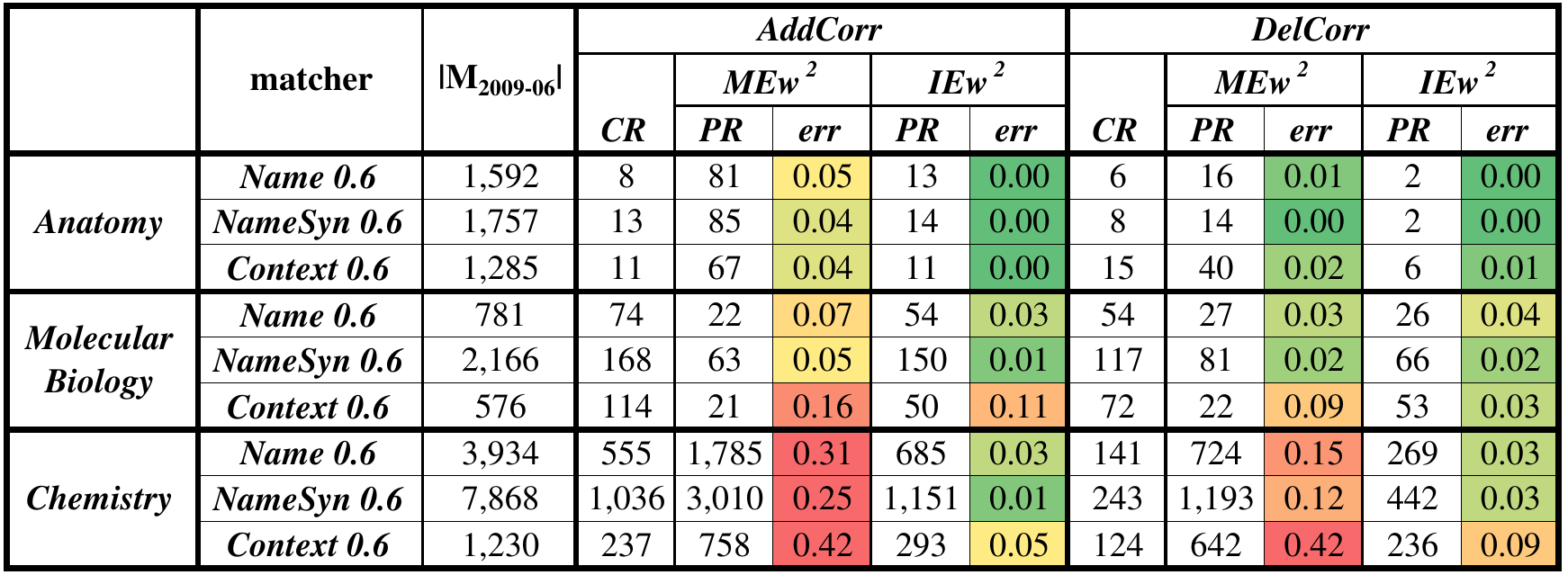}
	\caption{Number of correct and estimated $AddCorr$ and $DelCorr$ operations using mapping versions $M_{2008-06}$-$M_{2008-12}$ to predict changes in $M_{2009-06}$. Comparison of two methods ($ME w\textsuperscript{2}$, $IE w\textsuperscript{2}$), for three life science domains and the three matchers. $CR$ ($PR$) - number of correct (predicted) result, $err$ - error rate on a red (high $err$) green (small $err$) scale}
	\label{fig:predictionMethods}
\end{figure}

To get an impression how many change operations we predict for \emph{ME w\textsuperscript{2}} and \emph{IE w\textsuperscript{2}}, we selected the following case. Considering the change factors in~Fig.\ref{fig:change_factors_allDomains} we would expect that it is hard to predict version 2009-06 based on 2008-06 and 2008-12 for all three match scenarios. In particular, for \textit{Anatomy} and \textit{Chemistry} we see a strong decrease in their change factors whereas for \textit{Molecular~Biology} an increase occurred. Fig.~\ref{fig:predictionMethods} shows detailed results of the prediction case. The error rate $err$ gives the absolute difference of $PR$ and $CR$ divided by the respective mapping size for the predicted version ($\left|M_{2009-06}\right|$). To get an better overview we illustrate $err$ on a red green scale. Overall both methods produce relatively good results (green $err$ values) for correspondence deletions. By contrast, estimating additions seems more complicated. \emph{IE w\textsuperscript{2}} produces only small errors for additions whereas \emph{ME w\textsuperscript{2}} either estimates too high (for \textit{Anatomy} and \textit{Chemistry}) or too low (for \textit{Molecular~Biology}) values (yellow to red $err$ values). This is triggered by the previous trend of mapping evolution, as we have seen in Fig.\ref{fig:change_factors_allDomains}. Thus, if the pattern of mapping evolution suddenly changes, methods making an estimation solely on the basis of previous mapping changes fail.

By contrast, $IE w\textsuperscript{2}$ involves knowledge about ontology evolution as well as its impact on mapping evolution which leads to more accurate prediction results. Especially considering the overall mapping sizes, the predicted results ($PR$) for \emph{Anatomy} are very close to the correct results ($CR$) (e.g., 8--13, 13--14, 11--11 for correspondences additions). In general, it seems very difficult to predict mapping changes for \emph{Chemistry} and the \emph{context 0.6} matcher. For \emph{Chemistry} one and the same ontology change factor can lead to mapping changes of different magnitude so that change prediction becomes a complex task. For \emph{context 0.6}, there are several different influences as the evolution of the concept itself, its parents and its children, making it difficult to correctly predict mapping changes.

In general we can recommend that the OAEI \emph{Anatomy} mapping is still feasible and reliable as there were relatively few ontology changes since 2007. Thus, we would expect only few mapping adaptations. By contrast, knowledge in the \emph{Molecular~Biology} or \emph{Chemistry} domains changed dramatically in the last 5 years. Thus, mapping adaptation is strongly recommended to obtain useful mappings.

\section{Related Work}
\label{sec:rel_work}
In the last decade, ontology matching to semi-automatically create ontology mappings has become an active research field (see ~\cite{ES07,R11} for  overviews). In the life sciences especially the matching of anatomy ontologies~\cite{ZB07} and molecular biological ontologies~\cite{BB05} has attracted considerable interest. Most match approaches focus on improving the quality of computed mappings by applying different matchers (e.g., based on the name/synonyms of concepts, the ontology structure or associated instances) in a workflow-like manner. For comparing available match systems w.r.t. their quality the OAEI~\cite{OAEI11} provides gold standard mappings, e.g., between MA and NCIT.

Previous work on ontology evolution (see~\cite{FMK+08,HTR11} for surveys) focused on ontology versioning~\cite{KFK+02}, the evolution process itself~\cite{S04} as well as the detection of changes between ontology versions~\cite{NM02}. Few approaches investigate how changes in ontologies should be propagated to dependent artifacts such as instances or annotations. For example, the ontology evolution process proposed in~\cite{SMM+02} includes a change propagation phase where performed changes are propagated to other ontologies that are based on the modified ontology. 

The evolution of ontology mappings has received only little attention so far. In our previous work~\cite{HKR08} we studied the evolution of ontologies, annotations and ontology mappings. We analyzed mapping evolution for one match problem and noticed dramatic increases in the number of correspondences especially for instance-based matchers. In a further study~\cite{THG+09} we focused on the stability of correspondences created by an instance-based matcher and proposed measures which allow for a classification of (un)stable correspondences.

In contrast to previous work this study focuses on the impact of ontology on mapping changes, i.e., we investigate (1) how ontology mappings change and (2) study how ontology changes correlate with mapping changes for different matchers. Furthermore, we use the knowledge from the correlation between ontology and mapping changes to estimate the cardinality of future mapping changes. The mapping versions under investigation were created with previously evaluated matchers such as name or name/synonym using the GOMMA system~\cite{KGHR11}. 
%We applied the COntoDiff algorithm~\cite{HGR10} to determine diffs between released ontology versions as basis for this enhanced study.

\section{Conclusion and Future Work}
\label{sec:summary}

We studied the evolution of ontology mappings and analyzed the ontology changes triggering mapping changes as well as the influence of different match techniques. Our analysis covered three life science mappings and three match strategies. Furthermore we proposed two prediction methods for estimating the cardinality of future mapping changes. Except for anatomy ontologies, we observed that ontology mappings based on common match strategies using name and synonym information often experience heavy changes.  Our prediction methods were quite effective and could reasonably estimate the number of correspondence additions and removals in a new mapping version. 
%Sometimes sudden changes in the ontology evolution trend occur. In these cases, the real impact of ontology changes on mapping changes helps to produce more accurate results. 
In future work, we plan to investigate how known ontology changes can be used to semi-automatically adapt ontology mappings without a completely new mapping determination.

% Based on a new highlevel classification of changes we determined the impact of ontology changes on ontology mapping changes. We were also interested how robust different metadata-based match strategies are w.r.t. ontology evolution. Moreover, we proposed generic prediction methods for estimating the cardinality of mapping changes depending on ontology changes. The goal/aim is to recommend users of ontologies whether or not they should adapt their ontology mappings. For instance, in case of heavy ontology extensions or revisions, it might be better to use the current ontology knowledge in the mappings. Conversely, an adaptation might not be necessary when only few ontology parts were changed meanwhile. To make predictions we introduce a simple mapping-based estimation method taking mapping changes from earlier versions into account. Beyond that our impact-based estimation method uses the concrete impact of ontology changes on mapping changes to better approximate the number of future correspondence additions or deletions in mappings.

\section*{Acknowledgment}
This work is supported by the German Research Foundation (DFG), grant RA 497/18-1 ("Evolution of Ontologies and Mappings").

\bibliographystyle{plain}
\bibliography{references}

\begin{thebibliography}{10}

\bibitem{AMA}
{Adult Mouse Anatomy}.
\newblock \url{http://www.informatics.jax.org/searches/AMA_form}.

\bibitem{BB05}
O.~Bodenreider and A.~Burgun.
\newblock Linking the gene ontology to other biological ontologies.
\newblock In {\em Proc. ISMB2005 SIG meeting on Bio-ontologies}, pages 17--18,
  2005.

\bibitem{BHR+05}
O.~Bodenreider, T.F. Hayamizu, M.~Ringwald, et~al.
\newblock Of mice and men: Aligning mouse and human anatomies.
\newblock In {\em Proc. of AMIA Annual Symposium}, 2005.

\bibitem{BS06}
O.~Bodenreider and R.~Stevens.
\newblock Bio-ontologies: current trends and future directions.
\newblock {\em Briefings in bioinformatics}, 7(3):256--274, 2006.

\bibitem{Chebi}
P.~De~Matos, R.~Alc{\'a}ntara, A.~Dekker, et~al.
\newblock Chemical entities of biological interest: an update.
\newblock {\em Nucleic acids res.}, 38(suppl 1):D249--D254, 2010.

\bibitem{DR02}
Hong-Hai Do and Erhard Rahm.
\newblock Coma: a system for flexible combination of schema matching
  approaches.
\newblock In {\em Proceedings of VLDB}, pages 610--621, 2002.

\bibitem{ES07}
J~Euzenat and P~Shvaiko.
\newblock {\em {Ontology matching}}.
\newblock Springer-Verlag New York, 2007.

\bibitem{FMK+08}
G.~Flouris, D.~Manakanatas, H.~Kondylakis, et~al.
\newblock Ontology change: Classification and survey.
\newblock {\em The Knowledge Engineering Review}, 23(2):117--152, 2008.

\bibitem{GO08}
{Gene Ontology Consortium}.
\newblock The gene ontology project in 2008.
\newblock {\em Nucleic Acids Res.}, 36(Database Issue):D440--D444, 2008.

\bibitem{GNM09}
A.~Ghazvinian, N.F. Noy, and M.A. Musen.
\newblock Creating mappings for ontologies in biomedicine: Simple methods work.
\newblock In {\em Proc. of AMIA Annual Symposium}, 2009.

\bibitem{GHKR11}
A.~Gross, M.~Hartung, T.~Kirsten, and E.~Rahm.
\newblock Mapping composition for matching large life science ontologies.
\newblock In {\em 2nd Intl. Conf. on Biomed. Ontology (ICBO)}, 2011.

\bibitem{HGR10}
M.~Hartung, A.~Gross, and E.~Rahm.
\newblock {Rule-based Generation of Diff Evolution Mappings between Ontology
  Versions}.
\newblock {\em CoRR}, abs/1010.0122, 2010.

\bibitem{HKR08}
M.~Hartung, T.~Kirsten, and E.~Rahm.
\newblock {Analyzing the evolution of life science ontologies and mappings}.
\newblock In {\em Data Integration in the Life Sciences}, pages 11--27, 2008.

\bibitem{HTR11}
Michael Hartung, James~F. Terwilliger, and Erhard Rahm.
\newblock Recent advances in schema and ontology evolution.
\newblock In {\em Schema Matching and Mapping}, pages 149--190. Springer, 2011.

\bibitem{JL05}
V.~Jakoniene and P.~Lambrix.
\newblock {Ontology-based integration for bioinformatics}.
\newblock In {\em VLDB Workshop on Ontologies-based techniques for DataBases
  and Information Systems-ODBIS 2005}, pages 55--58, 2005.

\bibitem{KGHR11}
T.~Kirsten, A.~Gross, M.~Hartung, and E.~Rahm.
\newblock Gomma: a component-based infrastructure for managing and analyzing
  life science ontologies and their evolution.
\newblock {\em Journal of Biomedical Semantics}, 2:6, 2011.

\bibitem{KFK+02}
M~Klein, D~Fensel, A~Kiryakov, and D~Ognyanov.
\newblock {Ontology versioning and change detection on the web}.
\newblock {\em Knowledge Engineering and Knowledge Management: Ontologies and
  the Semantic Web}, pages 247--259, 2002.

\bibitem{L07}
P~Lambrix, H~Tan, V~Jakoniene, and L~Strömbäck.
\newblock Biological ontologies.
\newblock In {\em Semantic Web: Revolutionizing Knowledge Discovery in the Life
  Sciences}, pages 85--99. Springer Verlag, 2007.

\bibitem{NCIT}
{NCI Thesaurus}.
\newblock \url{http://ncit.nci.nih.gov/}.

\bibitem{NM02}
N~F Noy and M~A Musen.
\newblock {Promptdiff: A fixed-point algorithm for comparing ontology
  versions}.
\newblock In {\em Proc. of Nat. Conf. on Artificial Intelligence}, pages
  744--750, 2002.

\bibitem{NSW+09}
N.F. Noy, N.H. Shah, P.L. Whetzel, et~al.
\newblock Bioportal: ontologies and integrated data resources at the click of a
  mouse.
\newblock {\em Nucleic acids res.}, 37(suppl 2):W170--W173, 2009.

\bibitem{OAEI11}
{Ontology Alignment Evaluation Initiative}.
\newblock \url{http://oaei.ontologymatching.org/}.

\bibitem{R11}
E.~Rahm.
\newblock {Towards Large Scale Schema and Ontology Matching}.
\newblock In {\em {Schema Matching and Mapping}}, chapter~1, pages 3--27.
  Springer, 2011.

\bibitem{RB01}
E~Rahm and P~A Bernstein.
\newblock {A survey of approaches to automatic schema matching}.
\newblock {\em The VLDB Journal}, 10(4):334--350, 2001.

\bibitem{S04}
L~Stojanovic.
\newblock {\em {Methods and tools for ontology evolution}}.
\newblock PhD thesis, University of Karlsruhe, 2004.

\bibitem{SMM+02}
L~Stojanovic, A~Maedche, B~Motik, and N~Stojanovic.
\newblock {User-driven ontology evolution management}.
\newblock {\em Knowledge Engineering and Knowledge Management: Ontologies and
  the Semantic Web}, pages 133--140, 2002.

\bibitem{TML07}
P.D. Thomas, H.~Mi, and S.~Lewis.
\newblock {Ontology annotation: mapping genomic regions to biological
  function}.
\newblock {\em Current opinion in chemical biology}, 11(1):4--11, 2007.

\bibitem{THG+09}
Andreas Thor, Michael Hartung, Anika Gross, Toralf Kirsten, and Erhard Rahm.
\newblock An evolution-based approach for assessing ontology mappings - a case
  study in the life sciences.
\newblock In {\em BTW}, pages 277--286, 2009.

\bibitem{Uberon}
{UBERON}.
\newblock \url{http://obofoundry.org/wiki/index.php/UBERON:Main_Page}.

\bibitem{ZB07}
S.~Zhang and O.~Bodenreider.
\newblock Experience in aligning anatomical ontologies.
\newblock {\em International journal on Semantic Web and information systems},
  3(2):1--26, 2007.

\end{thebibliography}

\end{document}